\PassOptionsToPackage{hyphens}{url}
\documentclass[
  reprint,
  amsmath,
  amssymb,
  aps,
  nofootinbib,
  longbibliography,
  floatfix
]{revtex4-2}

\usepackage[T1]{fontenc}
\usepackage[utf8]{inputenc}
\usepackage{graphicx}
\usepackage{xcolor}
\usepackage[colorlinks=true,citecolor=blue,linkcolor=blue,urlcolor=blue]{hyperref}
\setcitestyle{numbers,square,comma}

\begin{document}

\title{The Tone of Awareness: Topic, Sentiment, and Toxicity Maps During Mental Health Month on TikTok}

\author{Henrique Ferraz de Arruda}
\affiliation{Institute for Biocomputation and Physics of Complex Systems (BIFI), University of Zaragoza, Zaragoza, Spain}
\affiliation{ARAID Foundation, Zaragoza, Spain}

\author{Andreia Sofia Teixeira}
\affiliation{BRAN Lab, Network Science Institute, Northeastern University London, London, UK}
\affiliation{Kent Medway Medical School, Canterbury, United Kingdom}
\affiliation{LASIGE, Faculdade de Ciências da Universidade de Lisboa, Lisboa, Portugal}

\author{Pranay Gundala Reddy}
\affiliation{Observatory on Social Media, Indiana University, Bloomington, IN, USA}

\author{Anindya Mondal}
\affiliation{Observatory on Social Media, Indiana University, Bloomington, IN, USA}

\author{Kleber Andrade Oliveira}
\affiliation{Social Dynamics Research Lab, Department of Psychology, University of Limerick, Limerick, Ireland}

\author{Filipi Nascimento Silva}
\email{filipi.nascimento@kellogg.northwestern.edu}
\affiliation{Observatory on Social Media, Indiana University, Bloomington, IN, USA}
\affiliation{CSSI - Kellogg School of Management, Northwestern University, IL, USA}

\date{\today}

\begin{abstract}
Despite raising concerns about the mental health effects associated with the usage of TikTok, little is known about how related content is framed by creators and received by audiences. We collect the content of 28,341 TikTok videos and 80,130 comments from Mental Health Awareness Month (May) in 2023 and 2024 via the TikTok Research API, and study how the \emph{tone of awareness} varies across topics and years. We characterize ``tone'' as the emotional and interpersonal framing of mental health discourse, operationalized through sentiment and toxicity measures. We extract topics from video text using BERTopic and log-odds keywords, then quantify topic-conditioned sentiment (XLM-T) and toxicity (Detoxify) separately for video transcriptions and comments. Sentiment captures the affective valence of content, while toxicity reflects the presence of harmful or abusive language. We find a stable set of recurring themes across years, spanning clinical conditions, emotional disclosure, self-care, and campaign-oriented content, with engagement highly skewed toward a small subset of topics. All sentiment and toxicity analyses are computed separately for video content and comments, allowing us to distinguish between content production and audience reception. Sentiment in videos is often negative for emotionally charged topics, while comments tend to shift toward more mixed or positive polarity, especially for suicide prevention. Toxicity is low in median overall, but exhibits longer-tailed outliers in comments than in videos that are more pronounced in comments and concentrated in specific topics (e.g., ``Duet'', ``Suicide Prevention'', and ``Psychisch''). Overall, our results provide a topic-level decomposition of mental health discourse on TikTok during awareness-month campaigns.
\end{abstract}

\maketitle

\section{Introduction} 
\textbf{Disclaimer:} This paper analyzes mental health-related content on TikTok, including content that references suicide.

Mental health conditions affect more than one billion people around the world~\cite{whoMental}, leading to immense disability and economic consequences. Increasingly, online spaces are where millions share personal experiences daily. Social media platforms are central to this discourse, functioning both as spaces for emotional venting that can foster peer support and as places where vulnerable users may be exposed to harmful or hostile content.

TikTok has become one of the dominant platforms on which mental health content is produced and discussed. The combination of short-format video with the ease of sharing and commenting can turn intimate and emotionally vulnerable content into highly performative disclosure that ends up tied to engagement metrics. This can have both a positive effect, in which people feel a shared vulnerability online, but it can also turn into heavy and dysregulated usage. Recent literature has highlighted \emph{problematic TikTok use} reporting consistent associations with a range of mental health outcomes (e.g., depressive and anxiety symptoms), while also stressing that the effects are heterogeneous and likely depend on individual vulnerability, content diets, and social context \cite{jain2025exploring,galanis2025association}.

Policy debates are now following these concerns in young and adolescent populations, such that Australia has implemented a nationwide social media ban for users under the age of 16~\cite{fardouly2025potential} and Denmark is planning to follow suit for users under 15 years of age~\cite{apnews2025denmark}.

At the same time, doing careful and reproducible research with TikTok data is challenging and not straightforward. The platform is shaped by a proprietary personalized recommendation system, and the content itself is inherently multimodal: speech, on-screen text, captions, audio, and comment threads. This creates a mismatch with the partial traces researchers can typically access. Even when relying on the TikTok Research API, recent audits document constraints that directly affect measurement and replicability, including access and sampling restrictions, shifting time windows, incomplete or inconsistent metadata, and failures to retrieve content that is publicly observable \cite{corso2024we}.

A second challenge is conceptual. The \emph{tone} of mental health discourse is not one-dimensional: awareness content can be supportive, informational, and stigma-reducing, while simultaneously attracting dismissive, hostile, or triggering responses. In computational social science, sentiment provides one proxy for affective valence; toxicity provides a different proxy for harmful or hostile language. Toxicity detection has become a common instrument for quantifying antagonism in online discourse, including in cross-country studies of political interaction where toxicity scores (and subdimensions such as insult or identity attack) are related to engagement and network structure \cite{falkenberg2024toxicity}. In this work, we treat sentiment and toxicity as complementary lenses rather than interchangeable ones: they capture different mechanisms, and they can diverge within the same topical space.

In this paper, we study Mental Health Awareness Month (May) as a recurring, platform-wide context where mental health discourse is especially visible. Using TikTok data from May 2023 and May 2024, we ask how the \emph{tone of awareness} is characterized between posts and comments when we look at (i) which topics dominate the conversation, and (ii) how sentiment and toxicity are distributed within those topics. 
By ``tone of awareness,'' we refer to the emotional and interpersonal nature of discourse, as determined by the sentiment and toxicity of posts and comments. Specifically, we aim to answer the following research questions: RQ1: Which topics dominate the mental health discourse during Mental Health Awareness Month?
RQ2: How are sentiment and toxicity distributed among these topics, and how do they differ between posts and their comments?
Methodologically, we first extract topics from the posts' text using BERTopic \cite{grootendorst2022bertopic}, and for each topic, we quantify sentiment with XLM-T \cite{barbieri2022xlmt} and toxicity using Detoxify-style scores (toxicity and subdimensions such as insult, threat, identity attack) \cite{hanu2020detoxify}. This topic-centered view is essential: rather than reporting a single platform-level ``positivity'' or ``toxicity'' number, we compute distributions within each topic and compare them across video transcribed content and their corresponding comments. 

The remainder of this manuscript is organized as follows. First, we elaborate on related work to our methodology. Next, we describe the data collection and preprocessing steps, including how we separate text from video posts (description, audio transcription, and hashtags) and comments, and how we address platform-specific constraints. Then, we present details of our method, including the topic modeling pipeline, keyword identification, comparisons, and the sentiment and toxicity models. In the results section, we present the main findings. We then discuss the results in terms of their implications and limitations. Finally, we conclude with directions for future work.

\section{Related Work}
Social media has been serving as a space for emotional disclosure, making it an ideal source of data for studying behavior during Mental Health Month. Previous studies have identified venting as a prevalent form of expression online that often fosters stronger group cohesion and peer support~\cite{jalonen2014social,krems2024venting}. Jalonen~(2014) argues that social media function as ``an arena for venting negative emotions,'' emphasizing how individually felt emotions can be shared, amplified, and transformed into collective emotional dynamics through online interaction~\cite{jalonen2014social}. A psychological study suggests that venting can serve social functions~\cite{krems2024venting}. For example, airing grievances to others can increase social bonding or sympathy, even if it does not necessarily alleviate the anger of the person doing the venting~\cite{krems2024venting}. Vermeulen et al. (2018) examined how adolescents use social media to express positive and negative emotions~\cite{vermeulen2018smiling}. They often use hashtags such as \#smiling and \#venting to express their moods and encourage responses from their peers. With TikTok data, Basch et al. (2022) analyzed 100 mental health-related videos that collectively received over 1 billion cumulative views, finding that videos receiving above-average engagement often attracted comments offering support or describing similar struggles~\cite{basch2022deconstructing}.

Due to platform constraints, accessing TikTok data for research is challenging. TikTok's content is filtered through a proprietary recommendation algorithm and delivered in a multimodal format (i.e., video, text, and audio). This creates a discrepancy with the partial data traces that researchers can obtain. Even after launching an official Research API, audits have revealed limitations. For instance, Entrena-Serrano
et al.~(2025) found that the API failed to retrieve metadata for approximately one out of every eight videos, including some from high-profile accounts, resulting in an incomplete dataset~\cite{entrena2025tiktok}. Similarly, Corso et al. documented sampling biases and inconsistencies in the API results in a systematic audit~\cite{corso2024we}. Beyond technical limitations, Ungless et al.~(2025) report that TikTok users with marginalized identities (e.g., LGBTQ+ users) often experience content censorship, affecting the visibility of their posts~\cite{ungless2025experiences}. 

These limitations shape the types of analyses that are possible and underscore the need for careful methodological design. Our study addresses these constraints by focusing on a clearly defined time period (Mental Health Awareness Month in 2023 and 2024), sampling content with public mental health-related terms, and applying conservative thresholds to mitigate the effects of API limitations (e.g., capping the number of comments per topic).

Previous studies have also examined toxicity and sentiment on TikTok, particularly in marginalized or emotionally charged contexts. For example, Tibau et al.~(2025) examined hate speech against LGBTQ+ communities on TikTok and found that toxicity and supportive counter-speech often coexist ~\cite{tibau2025prevalence}. Avalle et al.~(2024) analyzed long-term engagement patterns across platforms and demonstrated that toxic interactions tend to follow recurring structures without reducing engagement~\cite{avalle2024platforms}. 

Other researchers assessed TikTok in the context of mental health~\cite{conte2025scrolling,mccashin2023using}. Zha and Chang (2026) proposed an analytical pipeline and showed that using multimodal data enhances the quality of TikTok data analysis~\cite{zha2026interpreting}. Another study examined mental health-related content and communities on TikTok from a user-centered perspective~\cite{milton2023see}. The study found that users' perceptions of belonging to a community, combined with TikTok's content presentation mechanisms, create an environment that supports mental health discourse and provides users with meaningful social support.

Building on this body of research, our study applies topic-centered sentiment and toxicity measurement to mental health content. Moreover, we compare the tone of videos with that of their replies. Specifically, we apply these approaches to a topic-centered analysis, measuring sentiment and toxicity at the level of recurring mental health themes, and contrasting the tone of video posts with that of their comments.

\section{TikTok Dataset}
\label{sec:data}

Using the TikTok Research API, we collected data on videos posted between fifteen days before and fifteen days after the Mental Health Month (May) of 2023 and 2024. Our dataset was obtained from a curated list of mental health-related terms and queried the TikTok Research API.
To reduce bias toward a single condition or campaign, we included not only campaign-related hashtags but also awareness-related hashtags and multiple terms related to disorders and well-being.
Specifically, we queried the API for posts containing any of the following terms (either in the text description or as a hashtag): \texttt{MentalHealthAwareness}, \texttt{anxiety}, \texttt{mentalhealth}, \texttt{depression}, \texttt{mentalhealthawarenessmonth}, \texttt{mentalhealthmatters}, \texttt{mentalilness}, \texttt{stress}, \texttt{suicide}, \texttt{adhd}, \texttt{burnout}, \texttt{trauma}, \texttt{suicideprevention}, and \texttt{emotionalwellbeing}. 
After retrieving the videos, we collect all associated comments using the same API\footnote{The license for the TikTok API can be found at: https://www.tiktok.com/legal/page/global/terms-of-service-research-api/en.}. For each video, we concatenate the description, audio transcription (provided by the API), and hashtags into a single textual document, while the comments are analyzed in their original form.

In summary, we collected a total of 28,341 video posts from 16,843 users. Furthermore, we collected comments on these videos as of the data collection date (comments between February and March 2025), resulting in a total of 80,130 comments.
Comments often happened after the period of collection considered for video posts. We will refer to each video post simply as \textit{post} henceforth. Each post features various metadata, including a textual description (often with hashtags), which we will use to further characterize each post in terms of topic of conversation, sentiment, and toxicity. We will also refer to the post's reception in the form of the text contained in the comments replying to that post.

\begin{figure*}[t]
  \centering
  \begin{minipage}[b]{0.48\textwidth}
    \centering
    \includegraphics[width=\linewidth]{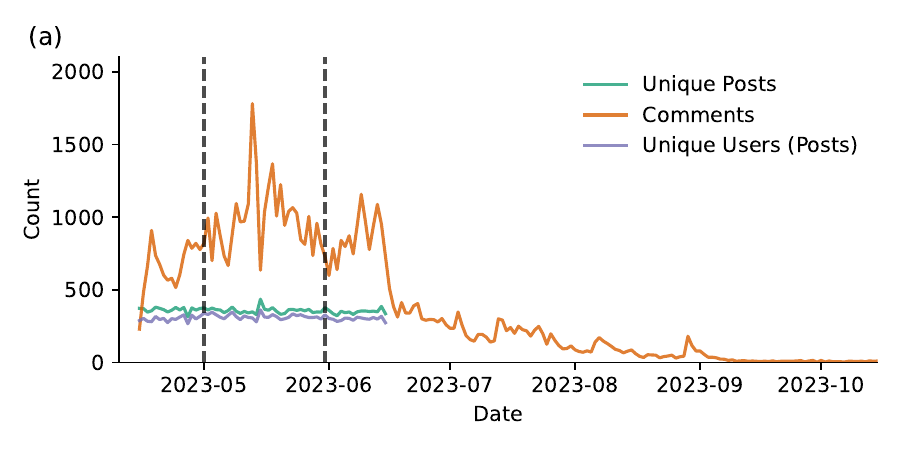}
  \end{minipage}
  \hfill
  \begin{minipage}[b]{0.48\textwidth}
    \centering
    \includegraphics[width=\linewidth]{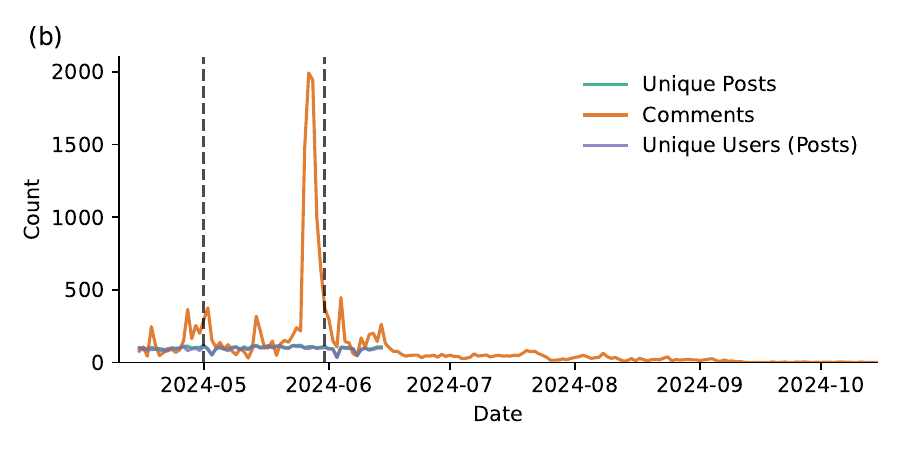}
  \end{minipage}
  \caption{Number of mental health-related posts and comments from the two periods: (a) 2023 and (b) 2024. The interval between the dashed lines marks Mental Health Awareness Month. We downloaded posts from the 15 days before and after Mental Health Awareness Month. For the comments, however, we downloaded all available data. We displayed data up to mid-October because the numbers tend to decrease in subsequent months.}
  \label{fig:counts}
\end{figure*}

Figure~\ref{fig:counts} shows the number of mental health–related TikTok posts and comments in 2023 and 2024. The interval between the dashed vertical lines marks Mental Health Awareness Month. 
We retrieved all the provided comments available in the API for the collected posts. However, we visualize comment counts only through mid-October because volumes tend to decrease in subsequent months, which would obscure earlier trends.

Figure~\ref{fig:counts}(b) shows a pronounced peak driven by a small number of highly commented posts. Specifically, five posts accounted for the highest comment counts during this period: 6,923, 1,483, 1,430, 921, and 622 comments, respectively. Since the TikTok API is expected to return at most 1,000 comments per post, but we have observed instances in which it returns more, we apply a conservative per-topic cap to avoid overrepresenting extreme engagement in downstream analyses (topic, sentiment, and toxicity). For topics exceeding 500 comments, we retain a uniformly random sample up to this limit. This threshold is based on the 2023 data, for which no topic exceeded this volume, preserving the full 2023 dataset while reducing the influence of outlier posts in 2024.

\section{Methodology}
\begin{figure}[ht]
  \centering
  \includegraphics[width=0.8\linewidth]{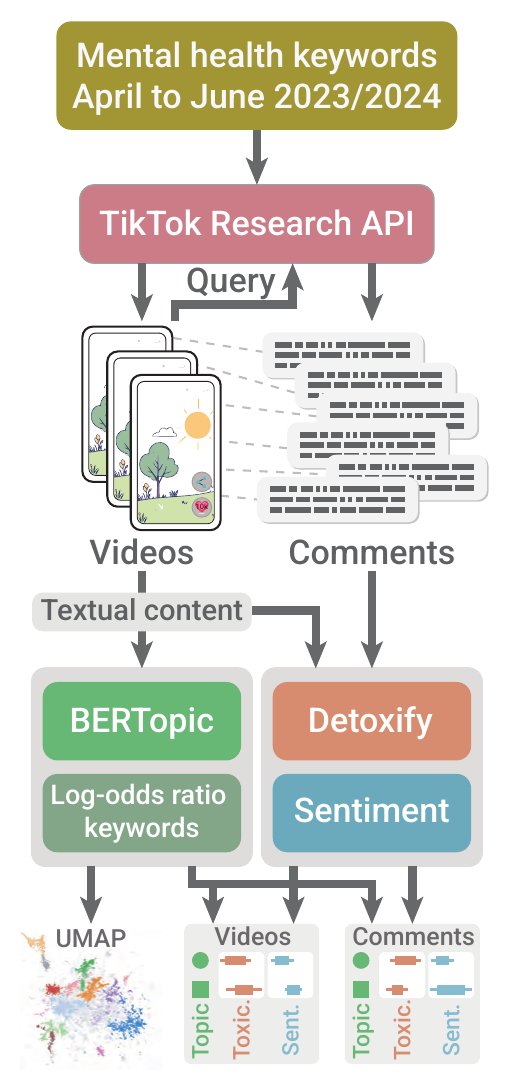}
    \caption{Schematic representation of our dataset collection and analysis pipeline.}
 
  \label{fig:schematic}
\end{figure}

We analyze mental health–related TikTok content using topic modeling, toxicity detection, and sentiment analysis to characterize both video narratives and audience reactions. Figure~\ref{fig:schematic} summarizes the full data collection and analysis pipeline\footnote{The codes necessary to reproduce this study can be found here: [To be added when published].}. 
We apply BERTopic\footnote{BERTopic is licensed under the MIT License (https://github.com/MaartenGr/BERTopic/blob/master/LICENSE).}~\cite{grootendorst2022bertopic} to identify thematic clusters in the textual content of videos. BERTopic is a topic modeling framework that combines transformer-based embeddings, dimensionality reduction, and density-based clustering. In particular, each document is first embedded using a sentence-transformer model, producing dense semantic representations. These embeddings are then projected into a low-dimensional space using UMAP\footnote{UMAP is licensed under BSD 3-Clause License (https://github.com/lmcinnes/umap/blob/master/LICENSE.txt).}~\cite{mcinnes2018umap} to preserve both local and global structure. Clustering is performed using HDBSCAN\footnote{HDBSCAN is licensed under OSI Approved (BSD) license. Here, we use the implementation of https://pypi.org/project/hdbscan/.}~\cite{mcinnes2017hdbscan, campello2013density}, which automatically finds the number of clusters and labels outliers. 
For each resulting cluster, BERTopic computes class-based TF-IDF (c-TF-IDF) scores, which highlight terms that are particularly representative of a topic relative to the entire corpus. These scores form the basis for identifying salient keywords and interpreting the semantic content of each topic. Appendix Table~\ref{tab:topic_model_hparams} summarizes the hyperparameter settings used in our BERTopic experiments.
We used LangDetect\footnote{ langdetect is licensed under the Apache License 2.0. PyPI. \url{https://pypi.org/project/langdetect/}
} to identify the language of each post, and found that more than 99\% were in English. Therefore, we decided to use an English model for BERTopic. Therefore, topic embeddings for non-English comments should be interpreted cautiously.

Although BERTopic provides native topic representations through class-based TF-IDF (c-TF-IDF), we use the \emph{fightin' words} approach~\cite{monroe2008fightin} as an additional post hoc feature-evaluation step to select the keywords used for manual topic labeling. This step is not used to estimate topics or to modify BERTopic's document assignments. Instead, after BERTopic assigns documents to topics, we compute log-odds ratios with an informative Dirichlet prior following \citet{monroe2008fightin}. For each topic \(t\), we compare the frequency of each word \(w\) among documents assigned to \(t\) with its frequency in the remaining corpus. The Dirichlet prior stabilizes estimates for sparse words and reduces the influence of rare, idiosyncratic terms. This choice reflects the distinction between topic representation and lexical feature evaluation. c-TF-IDF is useful for producing default BERTopic labels, but our goal is to identify terms that are distinctively associated with a given topic relative to the rest of the corpus. The log-odds approach is designed for this comparative task and accounts for sampling variation through Bayesian shrinkage. The resulting keywords are used only for qualitative validation and manual topic labeling; they do not affect the BERTopic embeddings, UMAP projection, HDBSCAN clustering, document-topic assignments, or downstream sentiment and toxicity analyses.

We measure textual toxicity using Detoxify\footnote{Detoxify is licensed under the Apache License 2.0 (https://github.com/unitaryai/detoxify/blob/master/LICENSE).}~\cite{hanu2020detoxify}, a BERT-based classifier trained on the Jigsaw toxic comment dataset. We use the multilingual variant of Detoxify, which is better suited to corpora that may contain some non-English text. Detoxify outputs probabilistic scores for multiple toxicity-related attributes, including \texttt{toxicity}, \texttt{severe toxicity}, \texttt{insult}, \texttt{threat}, and \texttt{obscene}. In our analysis, we focus primarily on the \texttt{toxicity} score. However, we also display scores for \texttt{obscene} and \texttt{insult}. The metric represents the probability that a given text contains harmful or abusive language. Since Detoxify produces continuous scores in $[0,1]$, we do not rely on discrete class labels. Instead, we analyze score distributions directly and compare them across videos and their associated comments.

For sentiment classification, we rely on XLM-T~\cite{barbieri2022xlmt}, a multilingual sentiment classifier built on top of XLM-RoBERTa~\cite{conneau2020xlmroberta}. XLM-T was further pre-trained on approximately 198 million tweets across multiple languages, making it well-suited for social media text. We use a publicly available implementation\footnote{XLM-T is licensed under the Apache License 2.0 (https://github.com/cardiffnlp/xlm-t/blob/main/LICENSE) and its implementation is available at \url{https://huggingface.co/cardiffnlp/twitter-xlm-roberta-base-sentiment}.} fine-tuned to classify sentiment into three categories: \texttt{positive}, \texttt{neutral}, and \texttt{negative}. For simplicity, we summarize sentiment using a \emph{polarity} score defined as the difference between the predicted probabilities of the positive and negative classes, i.e., $\mathrm{polarity} = p(\mathrm{positive}) - p(\mathrm{negative})$.
This results in a continuous measure in $[-1,1]$, where positive values indicate overall positive sentiment and negative values indicate negative sentiment. As with toxicity, we analyze sentiment polarity distribution across videos and comments to compare emotional tone between content creators and audience responses. We performed a limited manual validation on a small stratified sample of posts and comments to sanity-check the sentiment and toxicity model outputs.

We selected these tools because they have been used in closely related social-media settings, while also recognizing that none is a ground-truth measure of TikTok mental-health discourse. BERTopic has been applied to TikTok and Twitter datasets, including studies that combined coherence-based selection with qualitative review of topic relevance \citep{lee2024identifying_tiktok_bertopic, egger2022topic_modeling_comparison_twitter, murthy2024ecigarette_tweets_bertopic}. XLM-T has been used for sentiment analysis of TikTok video descriptions and manually checked against human annotations in prior TikTok research \citep{yang2025behavioral_addiction_tiktok}. Detoxify has been used in social-media toxicity studies and in TikTok comment auditing \citep{xue2025youth_safety_tiktok}. 

To formally assess whether topic-specific patterns differ from the overall platform baseline, we employ a non-parametric distributional test. For each topic $t$, we compare the distribution of each metric (toxicity and sentiment) against the corresponding distribution computed over all remaining data excluding topic $t$. This is done separately for videos and comments. Specifically, we used the Mann-Whitney $U$ test~\cite{hollander2013nonparametric} to evaluate if the topic-conditioned distribution significantly differs from the background distribution, without assuming normality. 

To control for multiple hypothesis testing across topics and metrics, we applied the Benjamini-Hochberg false discovery rate (FDR) correction~\cite{benjamini1995controlling} and report adjusted $q$-values instead of raw $p$-values.

To run our experiments, we used a cluster equipped with 32-core CPUs and NVIDIA H100 GPUs. These resources were required to accelerate the BERTopic topic modeling and toxicity inference pipelines.

\section{Results}
Using BERTopic, we identify a set of coherent thematic clusters from the combined video text and hashtags. Instead of relying solely on the internal c-TF-IDF rankings, we further characterize each cluster through log-odds scores (\emph{fightin' words} approach), allowing us to identify words that are statistically distinctive relative to the rest of the corpus. Based on these keywords, we manually assign interpretable topic labels.

Table~\ref{tab:topics} reports the resulting topics, including their identifiers, assigned names, and the most distinctive tags. The topics capture a diverse range of mental health discourse, spanning clinical conditions (e.g., anxiety, bipolar disorder, borderline personality disorder), coping strategies and emotional support (e.g., self-care, positivity), and event-driven content (e.g., Mental Health Awareness Month and suicide prevention).

\begin{table*}[htbp]
\caption{Topics found by our clustering of the text and hashtags from video descriptions using log odds on probabilities given by BERTopic.}
\label{tab:topics}

\resizebox{\textwidth}{!}{%
\begin{tabular}{rll}
\hline
\textbf{ID} & \textbf{Topic label} & \textbf{Top tags (by log-odds)} \\ \hline
1 & Sunset & sunset, relatable, camhs, tiktokph, crazybatgirlcrew, weragainstbullies \\ \hline
2 & Depression & depressionanxiety, depression, realateable, real\_realateable, realateable\_depression, real\_realateable\_depression \\ \hline
3 & Mental Health Month & um, month, like, know, mental, awareness \\ \hline
4 & Venting 1 & vent, vent\_vent, ventaccount, ventpost, vent\_relatable, relatable \\ \hline
5 & Mental Health & mental, health, mentalhealthsupport, mental\_health, mentalhealthrecovery, health\_matter \\ \hline
6 & Bipolar & bipolar, ptsd, bipolardisorder, ocd, ptsdawareness, trauma \\ \hline
7 & Anxiety & anxiety, anxietyrelief, anxiety\_anxietyrelief, panic, anxiety\_anxiety, anxious \\ \hline
8 & Self Care & selfcare, selflove, mindfulness, selfcare\_selflove, depression\_selfcare, anxiety\_depression\_selfcare \\ \hline
9 & Positivity & motivation, smile, onthisday, makeup, positivity, dayrunchallenge \\ \hline
10 & Grief & miss, dm, grief, dm\_open, griefjourney, reminder \\ \hline
11 & Borderline & bpd, adhd, borderlinepersonalitydisorder, autism, borderlinepersonalitydisorder\_bpd, bpd\_anxiety \\ \hline
12 & Duet & duet, duet\_user, user, duetcrashers, iamstillhere, isurvived \\ \hline
13 & Suicide Prevention & suicideprevention, suicideawareness, suicidepreventionmonth, suicideawarenessmonth, suicidepreventionline, suicideawarenessmonth\_suicideawareness \\ \hline
14 & Psychisch & ich, und, ptbs, kptbs, kptbs\_ptbs, psychischeerkrankung \\ \hline
15 & Venting 2 & vent, depression\_vent, vent\_depression, ventaccount, vent\_anxiety, vent\_depressionanxiety \\ \hline
\end{tabular}%
}
\end{table*}

Three topic representations are worth highlighting and clarifying. First, Topic 12 (\emph{Duet}) reflects the duet feature on TikTok, where users record a side-by-side response to an existing video. In our data, this topic is largely driven by reactions to a particular motivational creator (e.g., posts referencing two distinct songs and the creator). Consequently, \emph{Duet} captures an interaction format and engagement pattern rather than a single mental health condition. Second, Topic 14 (\emph{Psychisch}) contains primarily German-language tags and text, including PTBS and KPTBS, which are common German abbreviations for Posttraumatische Belastungsstörung (i.e., post-traumatic stress disorder -- PTSD) and Komplexe Posttraumatische Belastungsstörung (complex PTSD). Therefore, in the results, we interpret Topic 14 as a discussion cluster related to PTSD in German. We treat it as a language-specific topic when comparing sentiment and toxicity patterns across topics. Third, two topics are highly related to venting. While both topics are centered on the ``vent'' tag, Topic 15 contains more condition-specific tags (e.g., ``vent\_depression'' and ``vent\_anxiety''), suggesting a closer alignment with clinically oriented venting than Topic 4.

\begin{figure}[ht]
  \centering
\includegraphics[width=\linewidth]{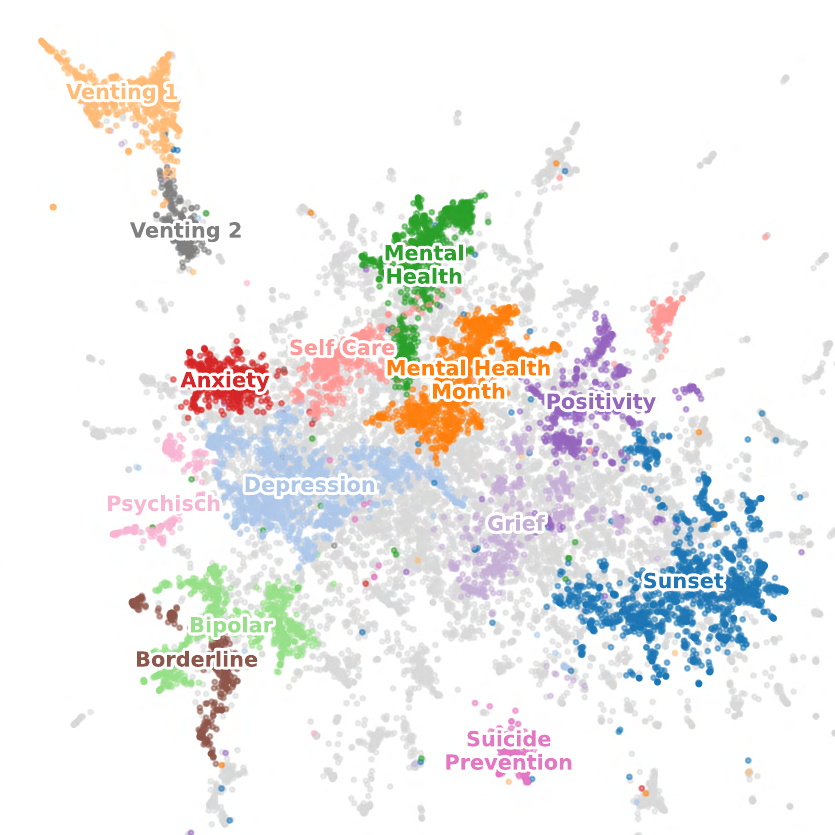}

  \caption{Zoomed-in region of the UMAP projection representing the topics of conversation identified in our clustering task. The topic named \emph{Duet} is not displayed because it is too far from the other clusters.}
  \label{fig:umap}
\end{figure}

Figure~\ref{fig:umap} presents the topic structure with a two-dimensional UMAP projection. This projection was generated using the UMAP coordinates produced by BERTopic, and the points were colored by their assigned topic. This visualization preserves the cluster structure identified by BERTopic, enabling us to visually inspect the separation and overlap among topics. This visualization excludes \emph{Duet} because its points lie far from the remaining clusters in the projection, and including it would substantially reduce the readability of the main topic structure. This also indicates that \emph{Duet} posts tend to differ the most from the others.

Topic 1 is named \textit{Sunset} because its most distinctive tag includes \textit{sunset}, mixed with tags associated with the TikTok ``sunrise or sunset'' challenge (see~\cite{distractify2025sunset}) and mental health-related tags. We do not interpret this topic as a clinically specific mental-health theme. Instead, it reflects a heterogeneous, platform-trend-driven cluster that entered the dataset because the trend co-occurred with mental health-related query terms or hashtags. This interpretation is consistent with its relatively dispersed position in the BERTopic embedding space. In contrast, more specific mental health themes form tighter groupings with clearer topical depth, such as \emph{Suicide Prevention} and \emph{Anxiety}.

\begin{figure}[t]
  \centering
  \includegraphics[width=\linewidth]{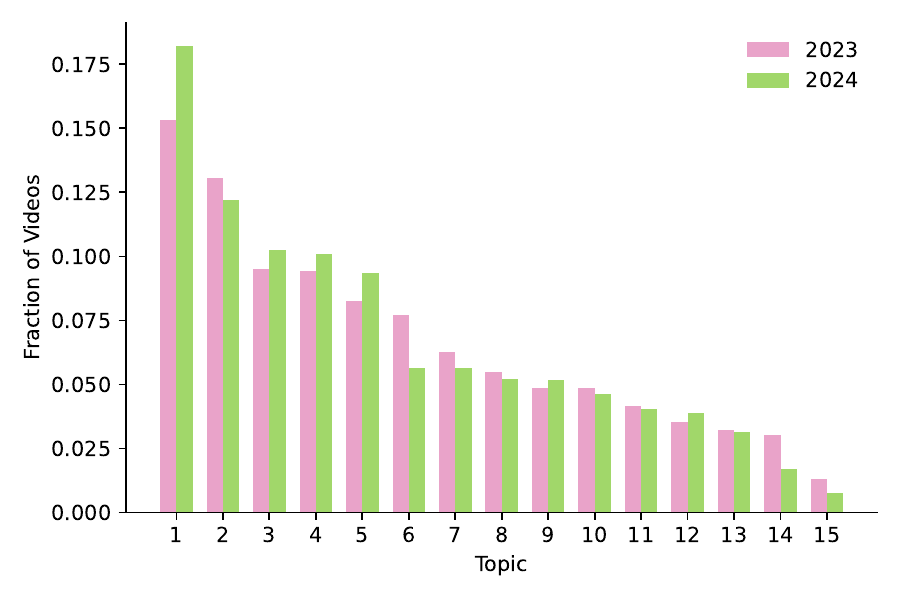}

  \caption{Fractions of posts per topic for the years 2023 and 2024.}
  \label{fig:counts_clusters}
\end{figure}

We now examine the composition of posts related to mental health in 2023 and 2024. Figure~\ref{fig:counts_clusters} compares the topics for each year, and it shows the percentage of posts assigned to each topic. Although the total number of collected posts varies between years, the relative distribution of topics remains stable, with most topics holding similar proportions in both years. This consistency suggests that our topic model captures recurring themes in mental health discourse on TikTok rather than artifacts specific to a given year, driven by dataset size or external yearly events. These results also support the use of the same set of topics to compare trends in sentiment and toxicity over time. 
Due to the stability of topic frequencies across years and the uncertainty surrounding the completeness of the TikTok data, we present the results of the analysis of sentiment and toxicity on the aggregated dataset rather than conducting a year-by-year comparison.

\begin{figure}[t]
  \centering
  \includegraphics[width=\linewidth]{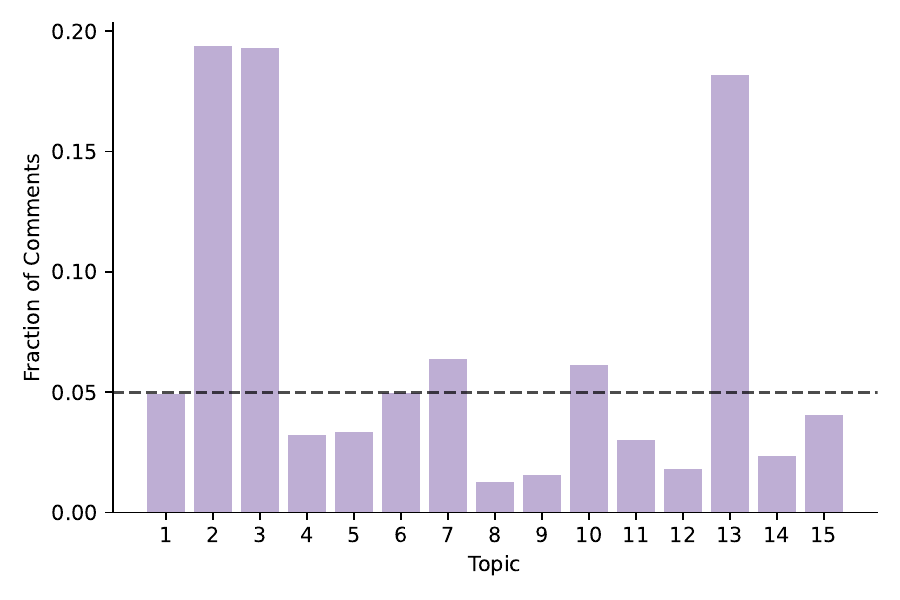}

  \caption{Fractions of comments by topic.}
  \label{fig:comments_counts_per_topic}
\end{figure}

Following this, in Figure~\ref{fig:comments_counts_per_topic} we present the fraction of comments associated with each topic. The distribution is strongly skewed; the top 5 most commented topics (Topics 2, 3, 13, 7, and 10) dominate comment activity, showing the largest engagement and reactivity with the prevalent mental burdens, such as \emph{Depression}, \emph{Anxiety}, \emph{Grief}, and \emph{Suicide Prevention}. However, as we will see later, sentiment and toxicity will vary among these topics. The remaining topics receive substantially fewer comments, indicating that the intensity of engagement follows a concentrated small subset of topics rather than being evenly distributed across all themes, which may be a result of less familiarity with either those mental disorders or practices (e.g., \emph{Self Care}).

\begin{figure*}[ht!]
\centering
\includegraphics[width=1\linewidth]{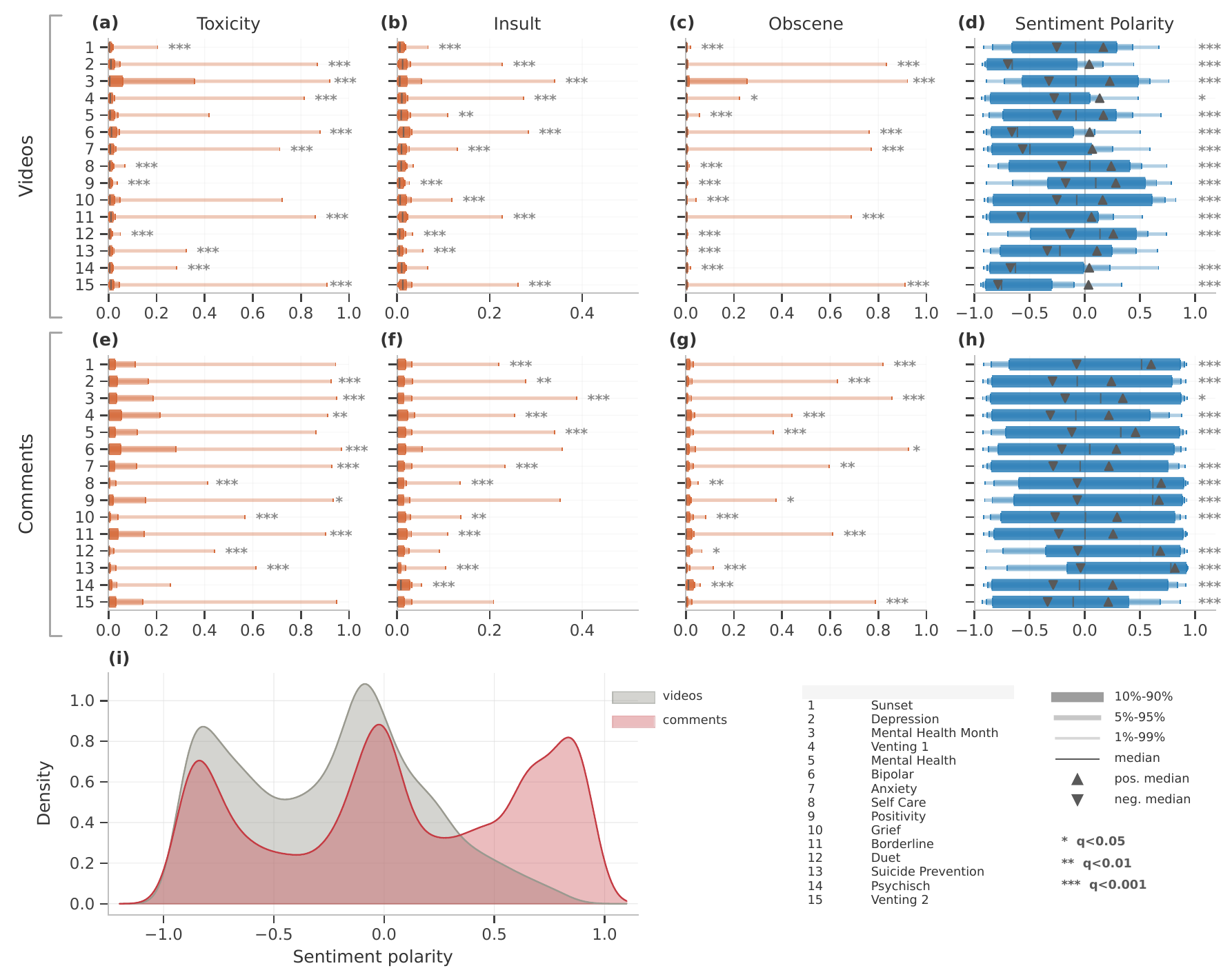}
  \caption{Toxicity and sentiment polarity results for posts and comments. Panels (a)-(c) display the distributions of Detoxify \textit{toxicity}, \textit{insult}, and overall \textit{obscene} scores for posts, and panels (e)-(g) the same metrics for comments. Panels (d) and (h) show the distribution of sentiment polarity for posts and comments, respectively. The polarity score is composed of the amount of tags labeled as ``positive'' (whose median is represented by a triangle) and those labeled as ``negative'' (median represented by an inverted triangle). Asterisks indicate statistically significant topic-level deviations from the background distribution, based on FDR-adjusted Mann–Whitney tests.  Panel (i) compares the kernel density estimates of polarity scores for posts and comments. }
  \label{fig:toxicity_sentiment_videos}
\end{figure*}

When looking into toxicity and sentiment, in Figure~\ref{fig:toxicity_sentiment_videos}, we report the topic-wise distribution of Detoxify scores (insult, obscene, severe toxicity, and overall toxicity) and sentiment polarity for each post's textual description across topics and the respective comments. To simplify our analysis, we omit the toxicity dimensions for which all observed scores are below 0.2, as these metrics show minimal variation. For each topic, we conducted Mann-Whitney $U$ tests comparing the distribution of each toxicity and sentiment metric within that topic against the remainder of the dataset (topic-vs-rest). Videos and comments were analyzed separately, resulting in $15$ topics $\times$ $4$ metrics $\times$ $2$ content types (videos and comments), for a total of $120$ statistical tests. To account for multiple comparisons, we applied the Benjamini-Hochberg false discovery rate (FDR) correction separately for each metric within each content type, resulting in $15$ adjusted comparisons per metric. Statistical significance is indicated using asterisk notation (given in terms of $q$-value). Bars without asterisks denote non-significant results after FDR adjustment. Focusing on posts (see Figures~\ref{fig:toxicity_sentiment_videos}~a-c), the medians of the toxicity-related scores are close to zero across topics, while the wider quantile bands (q01–q99) reveal a long-tailed pattern in which toxicity is concentrated in a relatively small number of outlier posts. Topic 3 (\emph{Mental Health Month}) stands out with comparatively larger upper-tail values across the toxicity dimensions. This indicates that the most extreme toxic posts are overrepresented in this topic relative to others. Topic 6 (\emph{Bipolar}) also exhibits toxic outliers, including in overall toxicity. This is consistent with the fact that a small fraction of posts contain highly toxic language, even though the median post remains non-toxic.

In contrast to toxicity, sentiment polarity patterns reflect the broader distribution of the data rather than primarily outliers (see Figure~\ref{fig:toxicity_sentiment_videos}~d). Clinical and venting-related topics, namely topics 2 (\emph{Depression}), 4 (\emph{Venting 1}), 6 (\emph{Bipolar}), 7 (\emph{Anxiety}), 11 (\emph{Borderline}), 14 (\textit{Psychisch}), and 15 (\emph{Venting 2}) skew negative in posting sentiment, whereas self-improvement and interaction-oriented topics, which are 8 (\emph{Self Care}), 9 (\emph{Positivity}), and 12 (\emph{Duet}) are more positive.

Figures~\ref{fig:toxicity_sentiment_videos}~(e)-(g) show the same metrics previously presented, for the comments. However, comment toxicity exhibits broader dispersion and heavier upper tails. This indicates that toxic language is more likely to appear in comments than in video descriptions. Notably, topic 14 (\emph{Psychisch}) shows substantially more toxic reception in comments than the corresponding post-level, with elevated upper-tail values across several toxicity dimensions.

Taken together, the contrast between videos and comments in sentiment and toxicity is not uniform across all mental health content but depends on the type of topic. Topics centered on clinical conditions, such as 6 (\emph{Bipolar}) and 11 (\emph{Borderline}), exhibit more pronounced toxicity outliers than support-oriented topics such as 8 (\emph{Self Care}) and 9 (\emph{Positivity}), whose toxicity distributions remain comparatively compressed near zero. Topic 11 (\emph{Borderline}) is primarily distinguished by both negative posting sentiment and pronounced toxicity outliers. This distinction is important for interpreting the ``awareness maps''. The main contrast is not simply that all condition-centered discourse is more toxic, but that some clinical topics combine negative sentiment with heavier toxic tails in reception, whereas self-care and positivity topics are characterized by more positive sentiment and less extreme toxicity.

Comment sentiment polarity is generally mixed, often spanning both negative and positive ranges within topics, but with clear topic-level differences (see Figure~\ref{fig:toxicity_sentiment_videos}~h). Topic 4 (\emph{Venting 1}) remains predominantly negative with slightly negative reception, while topics such as 8 (\emph{Self Care}), 9 (\emph{Positivity}), and 12 (\emph{Duet}) show more positive posting and reception. Finally, Topic 13 (\emph{Suicide Prevention}) exhibits one of the largest shifts between posting sentiment and reception, moving from comparatively negative polarity in posts to markedly more positive polarity in comments, consistent with supportive engagement in the responses.

Figure~\ref{fig:toxicity_sentiment_videos}~(i) complements the topic-level summaries in Figures~\ref{fig:toxicity_sentiment_videos}~(d) and~\ref{fig:toxicity_sentiment_videos}~(h) by showing the \emph{overall} distribution of sentiment polarity scores for posts versus comments. The kernel density estimates reveal a consistent shift toward more positive polarity in comments: comment distributions are centered farther to the right (higher polarity) than post distributions, which are more concentrated around neutral-to-negative values. This aggregate pattern is consistent with our per-topic results. While several topics contain predominantly negative sentiment in posts, comments often have a more positive or mixed reception. Together, the figure panels suggest that, even when the posted content is negative or distress-oriented, audience response is often positive. This difference is statistically significant based on FDR-adjusted Mann–Whitney tests ($q < 0.01$).

\section{Discussion}
Our results show that the mental health discourse on TikTok during Mental Health Awareness Month revolves around a consistent set of topics, including clinical conditions, emotional outlets, self-care, and public campaigns. The stability observed in 2023 and 2024 suggests that users have an enduring interest in this subset of themes. 

Furthermore, we observed a consistent asymmetry between the emotional tone of posts and comments. Many topics, especially those involving depression, anxiety, and venting, skewed negative in post sentiment, while the corresponding comment sections tended to be more positive or mixed. This pattern aligns with the findings of~\citep{krems2024venting}, who demonstrated that venting often elicits supportive responses from individuals who can relate to the distressed person. The strongest example in our data is Topic 13 (\emph{Suicide Prevention}), where negative posts are met with positive comments, suggesting community support.

Another interesting outcome of our analysis is the nuanced reaction to similarly emotionally charged topics. In the top 5 most commented topics, we have \emph{Grief} and \emph{Suicide Prevention}, and while the second demonstrated a shift in sentiment from video content to comments (from more negative to positive reflecting support), \emph{Grief} did not show that same pattern but a more neutral shift towards neutrality, suggesting a more neutral and less positive reaction.

Toxicity levels are generally low across both posts and comments, but outliers exist and are concentrated in a few topics. Comments tend to contain more toxic outliers than posts, consistent with prior work showing that, while rare, toxic replies have a disproportionate social impact~\cite{lee2025americans}. Conversely, we find that supportive comments are more prevalent than positive posts, which reinforces the notion that audience responses can mitigate the emotional impact of the original content. 
This heterogeneity matters for platform governance and when thinking about interventions. Blanket interventions risk treating all mental health discourse as equally charged, whereas our results suggest that potential harm is more concentrated in comment threads than in video narratives and varies across topics.

In summary, our results suggest that, although most posts and video content are not toxic, comment threads exhibit more negative sentiments and toxic outliers. These findings highlight comment sections as potential sources of harm and suggest that platform interventions, such as targeted moderation, friction, or support prompts, may be especially important in these contexts. Notably, Topic 13 (Suicide Prevention) shows one of the clearest post-comment shifts: posts are comparatively negative, while comments are more positive, suggesting supportive engagement in response to sensitive or distress-oriented content. At the same time, because suicide-related discourse is highly sensitive, even low-frequency toxic or dismissive replies can be harmful and require careful interpretation.

These findings are consistent with previous studies showing that emotional disclosure and support on social media can cause a variety of responses from the audience. In sensitive areas such as mental health, supportive, critical, and dismissive reactions tend to coexist. Our results extend this line of research by demonstrating that these dynamics are not uniform across content types, but rather, are concentrated in comment sections and vary across mental health topics.

\subsection{Limitations}
First, there are known constraints in the TikTok Research API. Previous studies have reported sampling gaps, missing metadata, and inconsistent coverage, all of which can impact the completeness and reproducibility of the data~\cite{entrena2025tiktok,corso2024we}. In our case, the API's limit on the number of comments per post restricted the quantity of available user responses, especially for popular posts. Additionally, the significant variation in text length between video descriptions and user comments may affect the classification accuracy for toxicity and sentiment polarity. While these limitations are important, they were carefully considered in the design and interpretation of our analysis. Where applicable, we adopted conservative thresholds and sampling strategies to mitigate their potential impact on the results. We notice that ~\cite{steel2025idsampling} developed a method to reverse-engineer TikTok video IDs and retrieve nearly 100\% of the activity at a given time. However, processing the data volume from this method may quickly become prohibitive in our case, since we are interested in the period of two months around the Mental Health Awareness Month, and they estimate that TikTok features about 269 million daily videos. A future effort to collect this full sample or other meaningful samples and contrast them to the TikTok Research API should address this important research gap, but it is out of our scope.

\section{Conclusion}
Given the impact of social media in everyday life, understanding engagement and the overall sentiment and toxicity of posts and comments is of utmost importance. With Depression being the major source of disability around the world \cite{friedrich2017depression} and with suicide being the the third leading cause of death among 15–29-year-olds \cite{whoSuicide}, it is crucial to understand how vulnerable populations, including young adults and individuals with lived experience, may be more susceptible to online content in both directions -- suggestions of self-care, but also negativity and judgment from those with less relatable experiences. With TikTok being highly used by younger populations, research studying its potential positive/negative effects is crucial, but far from being streamlined. Due to API restrictions, the sample of posts is not guaranteed to be representative of all genders, cultures, and languages, limiting the generalization of our results. Overall, our work contributes to a better understanding of how mental health-related topics are approached on TikTok. Despite API limitations, we were still able to deconstruct the dynamics of the online engagement in terms of mental health-related content distribution and corresponding reactions. We reveal how toxicity and sentiment shift from one to the other, and how emotionally charged topics may have different shifts, showing different levels of positivity and online support. 

Future research could examine characteristics beyond the topics that may attract higher toxicity or negative sentiment in the comments, such as the creator's characteristics or the post's style. A longitudinal analysis of toxicity dynamics could reveal periods when harmful responses peak. Additionally, since our results are restricted to a specific scenario and platform, another possibility for future work is to extend the same analysis to different social media platforms and subjects to measure whether the same differences in toxicity and sentiment exist between posts and comments.

\subsection{Ethical Considerations}
Due to the Data User Agreement\footnote{Available at \url{https://www.tiktok.com/legal/page/global/terms-of-service-research-api/en}.}, no data will be openly shared. Thus, potential ethical risks identified in studies like this (e.g., re-identification, misuse, etc.) are mitigated by not sharing either raw data or any direct quotes. Additionally, our study does not pinpoint causes or consequences of the content analyzed, and does not suggest any type of interventions.

\begin{acknowledgments}
HFA acknowledges ARAID for its financial support. KAO was partially supported by the European Union through ERC grant (ID-COMPRESSION, grant number: 101124175). Views and opinions expressed are however those of the author(s) only and do not necessarily reflect those of the European Union or the European Research Council Executive Agency. Neither the European Union nor the granting authority can be held responsible for them. AST acknowledges support by FCT through the LASIGE Research Unit, ref. UID/000408/2025. This work utilized Indiana University Jetstream2 CPU through allocation CIS230183 from the Advanced Cyber-infrastructure Coordination Ecosystem: Services \& Support (ACCESS) program, which is supported by National Science Foundation grants \#2138259, \#2138286, \#2138307, \#2137603, and \#2138296.
\end{acknowledgments}

\section*{Code and Data Availability}

The analysis code is available at \url{https://github.com/filipinascimento/tiktok-mental-health-analysis}. The shareable reindexed metrics dataset is archived on Zenodo at \url{https://doi.org/10.5281/zenodo.20646752}. This data release includes only topic labels, dates, video--comment linkage indices, and derived sentiment/toxicity scores; raw TikTok text, user identifiers, video identifiers, and comment identifiers are not redistributed.

\bibliographystyle{unsrtnat}
\bibliography{library}

\newpage
\appendix
\renewcommand{\thefigure}{A\arabic{figure}}
\renewcommand{\theHfigure}{appendix.figure.\arabic{figure}}
\setcounter{figure}{0}

\renewcommand{\thetable}{A\arabic{table}}
\renewcommand{\theHtable}{appendix.table.\arabic{table}}
\setcounter{table}{0}

\section{Hyperparameters used for topic modeling}

\begin{table}[htbp]
\centering
\small
\begin{tabular}{ll}
\hline
\multicolumn{2}{c}{\textbf{Sentence Embeddings}} \\
\hline
Embedding model & \texttt{all-mpnet-base-v2} \\
\hline
\multicolumn{2}{c}{\textbf{UMAP}} \\
\hline
Neighbors (k) & 20 \\
Embedding dim. & 10 (2 for visualization) \\
Min. distance & 0.05 \\
Training epochs & 50,000 \\
Distance metric & cosine \\
\hline
\multicolumn{2}{c}{\textbf{HDBSCAN}} \\
\hline
Min. cluster size & 300 \\
Max. cluster size & 5,000 \\
Min. samples & 15 \\
Distance metric & euclidean \\
Cluster selection & Excess of Mass \\
\hline
\end{tabular}
\caption{Hyperparameters used for topic modeling.}
\label{tab:topic_model_hparams}
\end{table}

\end{document}